\documentclass[3p,times,twocolumn]{elsarticle}

\usepackage{ecrc}

\volume{00}

\firstpage{1}

\journalname{Nuclear Physics B Proceedings Supplement}

\runauth{}

\jid{nuphbp}

\jnltitlelogo{Nuclear Physics B Proceedings Supplement}

\usepackage{amssymb}

\usepackage[figuresright]{rotating}

\begin{document}

\begin{frontmatter}



\dochead{}

\title{Small scales structures and neutrino masses}


\author{Francisco Villaescusa-Navarro}

\address{INAF - Osservatorio Astronomico di Trieste, Via Tiepolo 11, 34143, Trieste, Italy\\
INFN sez. Trieste, Via Valerio 2, 34127 Trieste, Italy}

\begin{abstract}
We review the impact of massive neutrinos on cosmological observables at the linear order. By means of N-body simulations we investigate the signatures left by neutrinos on the fully non-linear regime. We present the effects induced by massive neutrinos on the matter power spectrum, the halo mass function and on the halo-matter bias in massive neutrino cosmologies. We also investigate the clustering of cosmic neutrinos within galaxy clusters.
\end{abstract}

\begin{keyword}
Massive neutrinos \sep Cosmology


\end{keyword}

\end{frontmatter}



\section{Introduction}
\label{Sec:introduction}

The Big Bang theory predicts the existence of a cosmic neutrino background (CNB). Since neutrinos decoupled from the primordial plasma when they were ultra-relativistic, their momenta will follow a Fermi-Dirac distribution with a temperature equal to $T_\nu(z)=(4/11)^{1/3}T_{\rm CMB}(z)$, where $T_{\rm CMB}(z)$ is the temperature of the cosmic microwave background (CMB). 

At the linear order, massive neutrinos impact on cosmology in two different ways \cite{Lesgourgues_Pastor}: modifying the matter-radiation equality time and slowing down the growth of matter perturbations. It can be shown that these two effects induce a suppression, on small scales, in the matter power spectrum with respect to a cosmology with massless neutrinos. 
The above effects are commonly used to put tight constrains on the neutrino masses from cosmological observables such as the anisotropies in the CMB and the spatial distribution of galaxies at low redshift.

While the signatures left by massive neutrinos at the linear level have allowed us to put tight constrains on their masses, a huge amount of information, contained in non-linear scales, is not used since their theoretical counterpart has not been fully investigated. Therefore, in order to extract the maximum information from the cosmological observables it is crucial to study the impact of massive neutrinos into the fully non-linear regime. There are different ways to carry this task out: by means of perturbation theory, semi-analytic models or N-body simulations. Among those, the most precise one is the use of N-body simulations, although it is also the most computationally expensive.

\section{Signatures in the non-linear regime}
\label{Sec:signatures}

We have run N-body simulations using the code {\sc GADGET-III} \cite{Gadget}; massive neutrinos are simulated as additional particles. Our simulation suite comprises three different 
cosmologies: $\sum m_\nu=0.0$ eV, $\sum m_\nu=0.3$ eV and $\sum m_\nu=0.6$ eV. We note that even thought models with $\sum m_\nu\gtrsim0.15$ eV are ruled out by the latests analysis that combine CMB, BAO and Lyman-$\alpha$ forest data \cite{Costanzi_2014, Palanque_2014}, our purpose is to investigate the impact of massive neutrinos into non-linear observables as a function of the neutrino masses; thus, we are justified to consider such cosmological models.

\begin{figure}
\begin{center}
\includegraphics[width=0.42\textwidth]{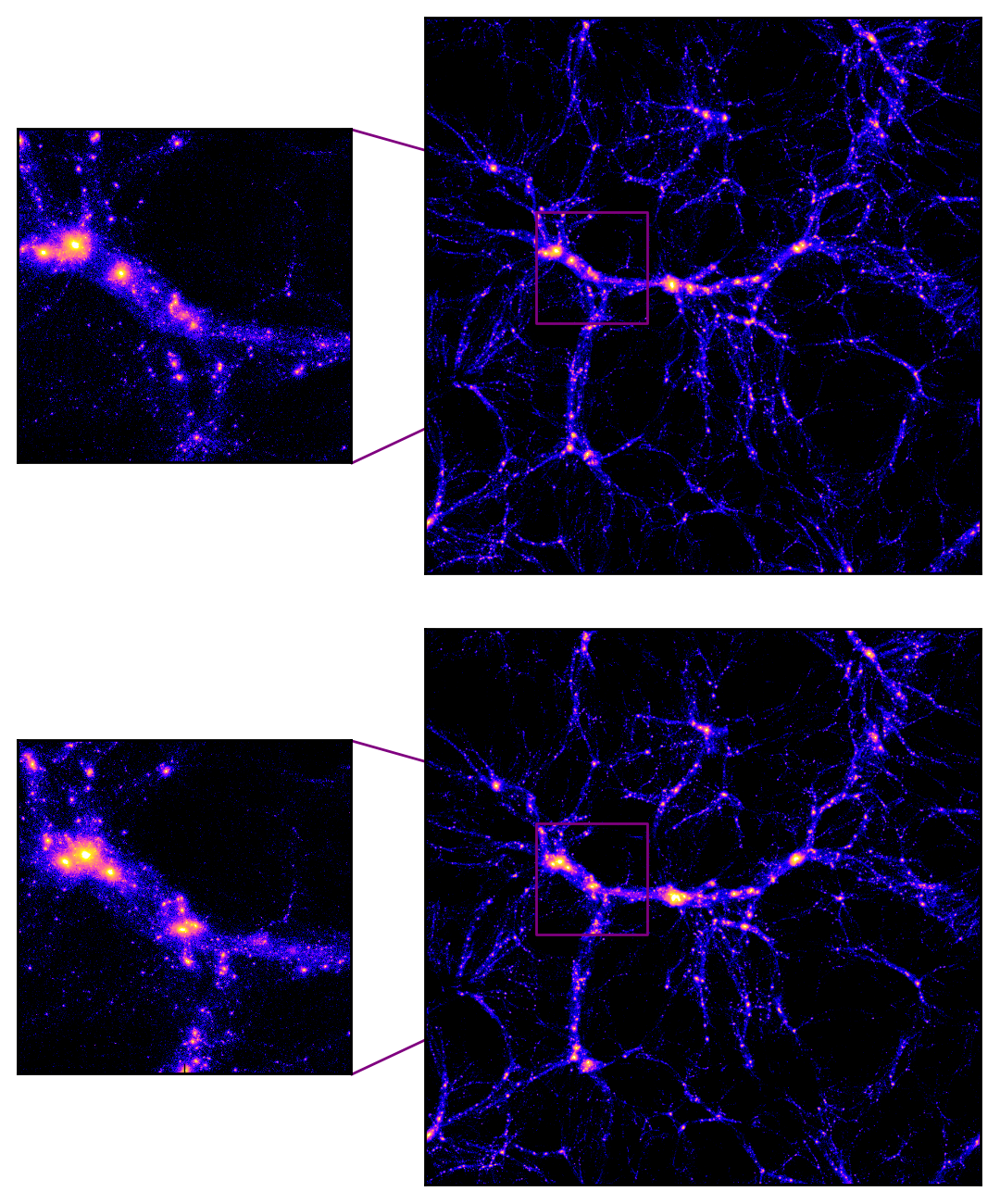}\\
\end{center}
\caption{Spatial distribution of the matter density field in a universe with $\sum m_\nu=0.6$ eV (top) and massless neutrinos (bottom) taken from a slice of 2 $h^{-1}$Mpc of our N-body simulations at $z=0$. The big panels show the distribution on large scales whereas the small panels display a zoom into the regions marked in purple.}
\label{matter_distribution}
\end{figure}

In Fig. \ref{matter_distribution} we show the total matter spatial distribution as obtained from our simulations, at $z=0$, for two different cosmologies: a model with $\sum m_\nu=0.0$ eV (top) and a model with $\sum m_\nu=0.6$ eV (bottom). the left panels show a zoom into the region marked with a purple square in the right panels. The differences among the two models can be seen directly from visual inspection. In particular, it is interesting to note that the model with massless neutrinos is a more "evolved" version of the model with $\sum m_\nu=0.6$ eV neutrinos. This can be seen from the bottom-left panel, where two halos are merging with the big halo located in the middle-upper part of that panel. However, for the model with massless neutrinos those two halos are further apart, indicating that the evolution in that universe is slower, as predicted by linear theory.

The fully non-linear matter power spectrum represents the most basic observable to investigate the impact of massive neutrinos. Several authors have already studied in detail the neutrino signatures in the matter power spectrum on small scales in the non-linear regime by means of N-body simulations \cite{Bird_2012, Brandbyge}. It is found that massive neutrinos suppress power on small scales, with respect to the equivalent massless neutrino cosmology. This suppression 
is larger than the linear prediction up to $k\sim1~h{\rm Mpc}^{-1}$, whereas on smaller scales the suppression is lower than the linear one. These features can be explained using an extension of the halo model as the one presented in \cite{Massara_2014}.

An important observable in which massive neutrinos leave their signature is in the halo mass function (HMF), i.e. the abundance of dark matter halos of a given mass. Theoretical studies have pointed out that the mass function becomes universal \cite{Sheth_Tormen} once it is expressed in terms of the peak height, $\nu=\delta_{\rm c}/\sigma(M)$, where $\delta_{\rm c}$ is the critical overdensity required for collapse ($\delta_{\rm c}\sim1.686$) and $\sigma(M)$ is the r.m.s. value of the linear density field when smoothed with a top-hat filter of mass $M$. It has been a common practice to estimate the HMF in massive neutrino cosmologies by computing $\sigma(M)$ using the total matter linear power spectrum. Recently, in a series of papers \cite{Villaescusa-Navarro_2014, Castorina_2014, Costanzi_2013} we have demonstrated using N-body simulations that the HMF in massive neutrinos cosmologies can be well reproduced if $\sigma(M)$ is computed with respect to the cold dark matter field, instead of the total matter field. Besides, by doing that, the HMF becomes universal. The use of an incorrect HMF for massive neutrinos cosmologies may bias the inferred values of the cosmological parameters as pointed out in \cite{Costanzi_2014}.

Massive neutrinos also imprint their signature on the bias between the spatial distribution of dark matter halos and the one of the underlying matter. It is well known that, for massless neutrinos cosmologies, on large, linear scales, the halo-matter bias should become scale-independent; we have confirmed this by using our massless neutrinos N-body simulations. However, in massive neutrinos cosmologies we find the halo-matter bias to be scale-dependent, even on very large scales. This can be understood if we account for the fact that halos trace the spatial distribution of cold dark matter, rather than all the matter. Our results point out that if the bias is defined as the square root of the ratio between the halos and cold dark matter power spectra, then the bias is scale-independent and universal.

\begin{figure}
\begin{center}
\includegraphics[width=0.48\textwidth]{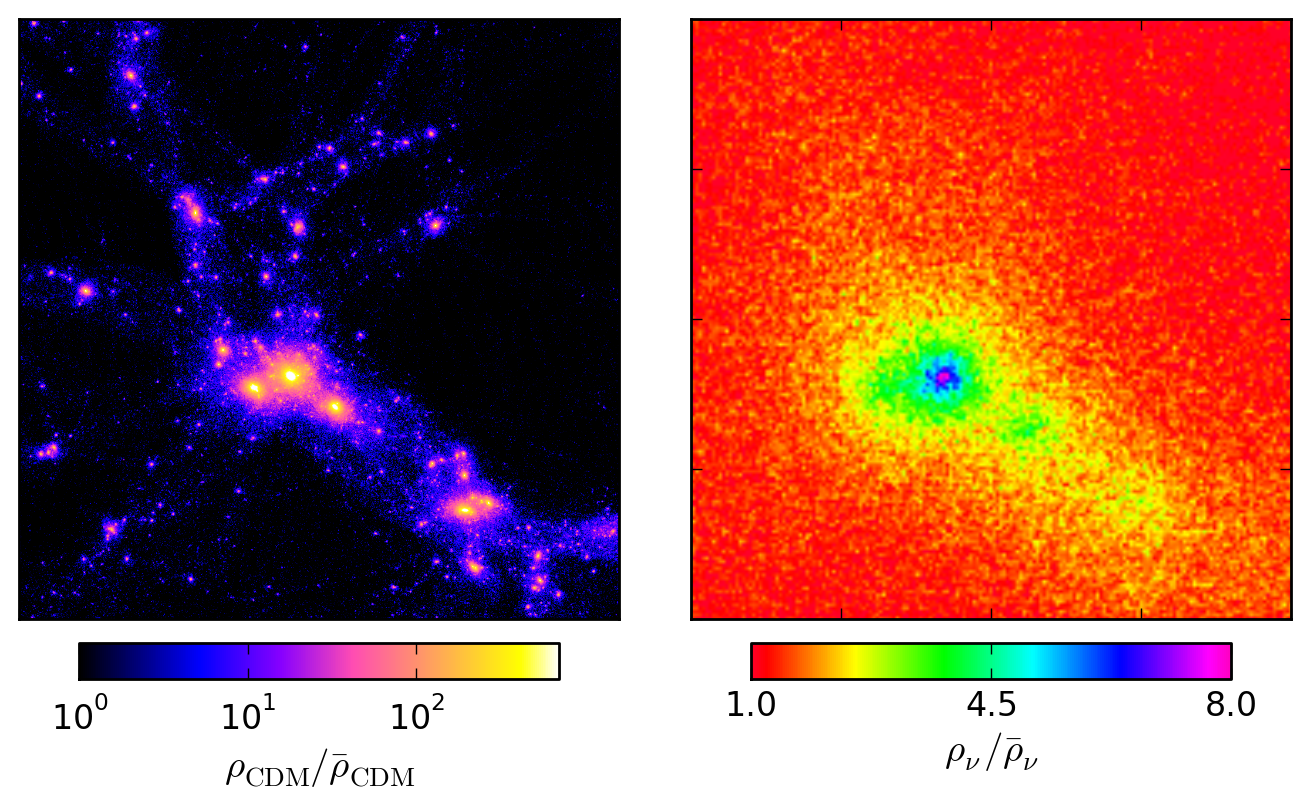}\\
\end{center}
\caption{Cold dark matter (left) and neutrino (right) overdensity around a massive dark matter halo at $z=0$ for a cosmological model with $\sum m_\nu=0.6$ eV neutrinos.}
\label{nu_clustering}
\end{figure}

Another interesting feature of the CNB is that some of the cosmic neutrinos may get trapped within the deep gravitational potential wells present in galaxy clusters. To see this more clearly let us write down some numbers. The velocity dispersion of galaxy clusters, at $z=0$, is of the order  $\sim 1000~{\rm km/s}$. On the other hand, the  mean thermal velocities of cosmic neutrinos is given by
$\bar{V}_\nu(z)=160\left(\frac{1+z}{m_\nu}\right)~{\rm km/s},$
where $m_\nu$ is the mass of each neutrino eigenstate in eV, i.e. $\bar{V}_\nu(z)\cong800~{\rm km/s}$ for neutrinos with $\sum m_\nu=0.6 ~{\rm eV}$ at $z=0$. Therefore, neutrinos in the low-velocity tail of the Fermi-Dirac distribution are expected to cluster within the most massive dark matter halos. This clustering of relic neutrinos can be seen in Fig. \ref{nu_clustering}. In the left panel of that figure we show the overdensity of CDM around one of the most massive halos in our simulation with $\sum m_\nu=0.6~{\rm eV}$ neutrinos, while the right panel instead displays the massive neutrinos overdensity. 

As can be seen from Fig. \ref{nu_clustering}, massive neutrinos do cluster around massive dark matter halos. There are few interesting features to note. The first one is that cosmic neutrinos do not cluster as much as the CDM. This can be seen from the scale located below the panels of Fig. \ref{nu_clustering}. The reason is that only neutrinos from the low-velocity tail of the momentum distribution can cluster. The second feature is that the \textit{neutrino halo} it is much more spatially extended that its CDM counterpart. This can be understood taking into account the large thermal velocities associated with neutrinos, which make them to have very large orbits. Moreover, those thermal velocities avoid the clustering of neutrinos in the center of the dark matter halo, inducing a core in the neutrino density profile. 

Authors in \cite{Villaescusa-Navarro_2013} studied the shape and amplitude of the neutrino density profile using N-body simulations. They found that the density profile of the neutrino halos can be well fitted by the expression 
$\delta_\nu(r)=(\rho_\nu(r)-\bar{\rho}_\nu)/\bar{\rho}_\nu=\rho_c/(1+(r/r_c)^\alpha),$
where $\rho_c$, $r_c$ and $\alpha$ are free parameters. Moreover, those authors demonstrated that that profile is universal, i.e. valid for all neutrino masses, for different host dark matter halo masses...etc.

\section{Summary and Conclusions}
\label{Sec:conclusions}

The existence of a cosmic neutrino background is predicted by our standard cosmological model. The dynamics of those relic neutrinos is very different to the one of the cold dark matter, since the thermal velocities of the former are very large whereas they are assumed to be negligible for the latter. The impact of massive neutrinos on cosmological observables is very well understood at the linear order, while the signatures left by neutrinos into the fully non-linear regime need sophisticated, and computationally expensive, tools as N-body simulations.

By running large box size N-body simulations that incorporate massive neutrinos as additional particles we have investigated the impact of massive neutrinos on: the matter power spectrum, the halo mass function, the halo-matter bias and the neutrino clustering within galaxy clusters. 

We find that the halo mass function in a massive neutrinos cosmology can be reproduced from the fitting functions of the corresponding massless neutrinos cosmology, by computing the relevant quantities (such as $\sigma(M)$) with respect to the CDM field, instead of the total matter field. By doing so we find that the halo mass function becomes universal. 

The halo-matter bias, defined as the square root of the ratio between the halos power spectrum to the total matter power spectrum, exhibits a scale-dependence behavior even on linear scales. This is a unique feature not present in standard massless neutrinos cosmologies. Our results point out that if we define the bias as the square root of the ratio between the halos and cold dark matter power spectra, then the bias becomes scale-independent on large scales and universal.

The momenta of the cosmic neutrino background follow a Fermi-Dirac distribution. Thus, neutrinos from the low velocity tail of the distribution are expected to cluster within the deep gravitational potential wells present in galaxy clusters. In our simulations we find halos of neutrinos located in the potential wells of massive dark matter halos. We find the density profile of the neutrino halos to be universal. Possible detection of such neutrino halos may be achieved through weak lensing \cite{Paco_2011}.

{\bf Acknowledgments.} FVN is supported by the ERC Starting Grant ``cosmoIGM'' and partially supported by INFN IS PD51 ``INDARK''.




\nocite{*}
\bibliographystyle{elsarticle-num}
\bibliography{Bibliography}

\begin{thebibliography}{10}
\expandafter\ifx\csname url\endcsname\relax
  \def\url#1{\texttt{#1}}\fi
\expandafter\ifx\csname urlprefix\endcsname\relax\def\urlprefix{URL }\fi
\expandafter\ifx\csname href\endcsname\relax
  \def\href#1#2{#2} \def\path#1{#1}\fi

\bibitem{Lesgourgues_Pastor}
J.~{Lesgourgues}, S.~{Pastor}, {Massive neutrinos and cosmology}, \physrep 429
  (2006) 307--379.

\bibitem{Gadget}
V.~{Springel}, {The cosmological simulation code GADGET-2}, \mnras 364 (2005)
  1105--1134.

\bibitem{Costanzi_2014}
M.~{Costanzi}, B.~{Sartoris}, M.~{Viel}, S.~{Borgani}, {Neutrino constraints:
  what large-scale structure and CMB data are telling us?}, \jcap 10 (2014) 81.

\bibitem{Palanque_2014}
N.~{Palanque-Delabrouille}, C.~{Y{\`e}che}, J.~{Lesgourgues}, G.~{Rossi},
  A.~{Borde}, M.~{Viel}, E.~{Aubourg}, D.~{Kirkby}, J.-M. {LeGoff}, J.~{Rich},
  N.~{Roe}, N.~P. {Ross}, D.~P. {Schneider}, D.~{Weinberg}, {Constraint on
  neutrino masses from SDSS-III/BOSS Ly$\alpha$ forest and other cosmological
  probes}, ArXiv e-prints\href {http://arxiv.org/abs/1410.7244}
  {\path{arXiv:1410.7244}}.

\bibitem{Bird_2012}
S.~{Bird}, M.~{Viel}, M.~G. {Haehnelt}, {Massive neutrinos and the non-linear
  matter power spectrum}, \mnras 420 (2012) 2551--2561.

\bibitem{Brandbyge}
J.~{Brandbyge}, S.~{Hannestad}, T.~{Haugb{\o}lle}, B.~{Thomsen}, {The effect of
  thermal neutrino motion on the non-linear cosmological matter power
  spectrum}, \jcap 8 (2008) 20.

\bibitem{Massara_2014}
E.~{Massara}, F.~{Villaescusa-Navarro}, M.~{Viel}, {The halo model in a massive
  neutrino cosmology}, ArXiv e-prints.

\bibitem{Sheth_Tormen}
R.~K. {Sheth}, G.~{Tormen}, {An excursion set model of hierarchical clustering:
  ellipsoidal collapse and the moving barrier}, \mnras 329 (2002) 61--75.

\bibitem{Villaescusa-Navarro_2014}
F.~{Villaescusa-Navarro}, F.~{Marulli}, M.~{Viel}, E.~{Branchini},
  E.~{Castorina}, E.~{Sefusatti}, S.~{Saito}, {Cosmology with massive neutrinos
  I: towards a realistic modeling of the relation between matter, haloes and
  galaxies}, \jcap 3 (2014) 11.

\bibitem{Castorina_2014}
E.~{Castorina}, E.~{Sefusatti}, R.~K. {Sheth}, F.~{Villaescusa-Navarro},
  M.~{Viel}, {Cosmology with massive neutrinos II: on the universality of the
  halo mass function and bias}, \jcap 2 (2014) 49.

\bibitem{Costanzi_2013}
M.~{Costanzi}, F.~{Villaescusa-Navarro}, M.~{Viel}, J.-Q. {Xia}, S.~{Borgani},
  E.~{Castorina}, E.~{Sefusatti}, {Cosmology with massive neutrinos III: the
  halo mass function and an application to galaxy clusters}, \jcap 12 (2013)
  12.

\bibitem{Villaescusa-Navarro_2013}
F.~{Villaescusa-Navarro}, S.~{Bird}, C.~{Pe{\~n}a-Garay}, M.~{Viel},
  {Non-linear evolution of the cosmic neutrino background}, Journal of
  Cosmology and Astroparticle Physics 3 (2013) 19.

\bibitem{Paco_2011}
F.~{Villaescusa-Navarro}, J.~{Miralda-Escud{\'e}}, C.~{Pe{\~n}a-Garay},
  V.~{Quilis}, {Neutrino halos in clusters of galaxies and their weak lensing
  signature}, \jcap 6 (2011) 27.

\end{thebibliography}







\end{document}